\documentclass{jfm}

\usepackage{enumitem} 
\usepackage[cal=cm]{mathalpha}
\usepackage{physics}
\usepackage{indentfirst} 
\usepackage[normalem]{ulem}

\usepackage{graphicx}
\usepackage{newtxtext}
\usepackage{newtxmath}
\usepackage{natbib}
\usepackage{hyperref}
\hypersetup{colorlinks=true, urlcolor=blue, citecolor= black}

\newcommand{\RomanNumeralCaps}[1]
\linenumbers

\title{Induced Diffusion of Internal Gravity Waves: Directionality and Role in Ocean Mixing}

\author{Yue Wu\aff{1}
  \corresp{\email{ywu.ocean@gmail.com}},
  \and Yulin Pan\aff{1}}

\affiliation{\aff{1}Naval Architecture and Marine Engineering, University of Michigan, Ann Arbor, MI, USA}

\begin{document}
\maketitle

\begin{abstract}
Induced diffusion (ID), an important mechanism of spectral energy transfer in the internal gravity wave (IGW) field, plays a significant role in driving turbulent dissipation in the ocean interior. 
In this study, we revisit the ID mechanism to elucidate its directionality and role in ocean mixing under varying IGW spectral forms, with particular attention to deviations from the standard Garrett-Munk (GM) spectrum.
The original interpretation of ID as an action diffusion process, as proposed by McComas \textit{et al.}, suggests that ID is inherently bidirectional, with its direction governed by the vertical-wavenumber spectral slope $\sigma$ of the IGW action spectrum, $n \propto m^\sigma$.
In contrast, by evaluating the wave kinetic equation, we reveal a more complete depiction of ID, comprising both diffusive and scale-separated transfers that are rooted in energy conservation within wave triads.
Although the action diffusion may reverse direction depending on the sign of $\sigma$ (i.e., between red and blue spectral cases), the combined ID transfer consistently leads to a forward energy cascade at the dissipation scale, thereby contributing positively to turbulent dissipation. This supports the viewpoint of ID as a dissipative mechanism in physical oceanography.
This study presents a physically grounded overview of ID and offers insights into the specific types of wave-wave interactions responsible for turbulent dissipation.
\end{abstract}



\section{Introduction}
\label{sec:intro}

Internal gravity waves (IGWs) are ubiquitous features of the ocean and are generated when stratified fluids are perturbed. The oceanic IGW field is primarily energized at large scales by atmospheric and tidal forcings and dissipated at small scales. Given the scale separation between forcing and dissipation, interscale energy transfer is crucial for sustaining an energy cascade across the IGW continuum. Mechanisms facilitating this interscale energy transfer include wave-wave interactions (e.g., \citealt{Hasselmann1966, Hasselmann1967}), wave-mean/eddy interactions (e.g., \citealt{Kafiabad2019, Dong2020, Dong2023, Savva2021, Delpech2024}), and bottom scattering (e.g., \citealt{Kunze2004}). Of these pathways, wave-wave interactions are regarded in many studies as the dominant process in the interior of the ocean \citep{Polzin2011, Polzin2014}.

The study of interscale energy transfer via wave-wave interactions was pioneered by McComas \textit{et al.} in a series of publications \citep{McComas1977a, McComas1977b, McComas1981a, McComas1981b}. These works posited that interscale energy transfer is dominated by three types of nonlocal interactions (i.e., wave triads that are scale-separated either in vertical wavenumber, frequency, or both), namely parametric subharmonic instability (PSI), elastic scattering (ES), and induced diffusion (ID). This framework laid the theoretical foundation for finescale parameterization to infer turbulent dissipation \citep{Henyey1986, Gregg1989, Polzin1995, Polzin2014} but has recently been shown to be incorrect due to the overlooked role of local interactions \citep{Dematteis2022, Wu2023b}.

\begin{figure}
\centering
\includegraphics[width=.15\textwidth]{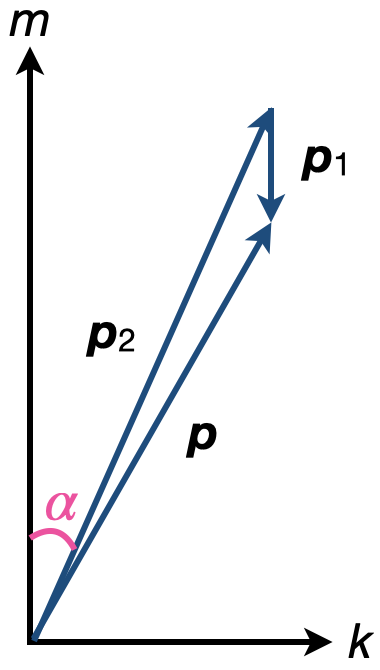}
\caption{A resonant wave triad $\mathbf{p}=\mathbf{p}_1+\mathbf{p}_2$ typical of the induced diffusion mechanism. 
The angle $\alpha$ between the wavenumber vector and the vertical is positively correlated with the wave frequency according to the dispersion relation.}
\label{fig:2}
\end{figure}

The present work focuses on the induced diffusion mechanism, one of the three types of nonlocal interactions. ID describes the scattering of a high-frequency, high-vertical-wavenumber wave $\mathbf{p}$ by a low-frequency, low-vertical-wavenumber wave $\mathbf{p}_1$, resulting in the generation of another high-frequency, high-vertical-wavenumber wave $\mathbf{p}_2$ through resonant interactions (FIG.~\ref{fig:2}). The dynamics at small scales (represented by $\mathbf{p}$ and $\mathbf{p}_2$) has been shown to satisfy a diffusion equation in terms of wave action $n$ (defined as wave energy $E$ divided by intrinsic frequency $\omega$), driving a diffusive cascade across vertical wavenumber $m$ \citep{McComas1977}
\begin{equation} \label{eq:ID}
\frac{\partial n(\mathbf{p})}{\partial t}=\frac{\partial}{\partial m}\left[D_{33}\frac{\partial}{\partial m}n(\mathbf{p})\right].
\end{equation}

Here, $\mathbf{p} = (k_x, k_y, m)$ denotes the three-dimensional (3D) wavenumber vector. Vertical diffusivity $D_{33}$, being the dominant component of the 3D diffusion tensor, is determined by the shear content of the largescale wave $\mathbf{p}_1$ (FIG.~\ref{fig:2})\footnote{An alternative perspective presented by \cite{Lanchon2023b} demonstrates that ID conserves the ratios $\omega/|m|$ and $k/m^2$ within a wave triad.}.
Assuming a stationary largescale field where $n(\mathbf{p}_1)$ and $D_{33}$ remain constant, \citet{McComas1981a} evaluated the downscale energy flux toward higher $m$ using \eqref{eq:ID}, assuming a logarithmic correction, $n \propto -\ln (m)$), at small scales to the standard Garrett-Munk (GM) spectrum. 
Turbulent dissipation is then approximated by the flux across the dissipation scale, where IGWs become unstable to shear instability and break at vertical scales smaller than 10 meters. As a result, ID was estimated to account for approximately $20\%$ of the total turbulent dissipation arising from all triad interactions, with the remainder attributed to PSI.

McComas's viewpoint on ID is not without problems under closer scrutiny. For the GM spectrum characterized by $n \propto m^0$ at small scales \citep{Cairns1976}, ID vanishes since the action spectrum displays no gradient in $m$. While secondary diffusion can arise from off-diagonal components in the diffusion tensor \citep{Dematteis2022}, the relative contribution of ID to the total turbulent dissipation is certainly much less than that postulated by McComas \textit{et al}. The situations for spectra deviating from GM are even more elusive, despite being commonly observed in field measurements \citep{Polzin2011}. Since diffusion always acts in the down-gradient direction, ID has the potential to reverse direction depending on the relative action intensity between the two smallscale waves, $\mathbf{p}$ and $\mathbf{p}_2$, in a single triad (FIG.~\ref{fig:2}). For an IGW spectrum, this direction is consequently governed by the sign of the vertical-wavenumber spectral slope, $\sigma$, of the action spectrum at small scales, $n \propto m^\sigma$. Specifically, for a blue or red action spectrum with positive or negative $\sigma$, the action diffusion at small scales corresponds to a forward or backward cascade, respectively. Does this imply that ID can contribute negatively to turbulent dissipation? This is a perplexing question, particularly given the long-standing consensus within the community that ID is a dissipative mechanism, as neither observational nor numerical evidence has reported a scenario involving a backward ID transfer (e.g., \citealt{Pan2020, Skitka2024a}). 

In this paper, we aim to resolve the confusion regarding the ID mechanism and develop a physically grounded understanding of its role in ocean mixing. Our study is based on the direct evaluation of the wave kinetic equation (WKE) that computes energy transfer arising from wave-wave interactions, along with a selection criterion to isolate the contribution of ID. 
We demonstrate that for GM-like spectra (with specific spectral forms summarized in Appendix~\ref{sec:AppA}), ID always contributes positively to turbulent dissipation. 
To reconcile this result with the reasoning regarding its reversible direction, one must recognize that ID is also associated with a scale-separated transfer, in addition to the diffusive transfer described by \eqref{eq:ID}. The scale-separated transfer (between the largescale wave, $\mathbf{p}_1$, and the two smallscale waves, $\mathbf{p}$ and $\mathbf{p}_2$; FIG.~\ref{fig:2}) is a direct consequence of energy conservation, as the diffusive transfer only conserves action but not energy. The energy flux across a chosen vertical scale is a combined result of both the scale-separated and diffusive transfers. For an action spectrum that is red or blue in $m$, the diffusive or scale-separated transfer dominates near the 10-meter vertical scale, respectively, leading to a consistently positive effect on ocean mixing. We conclude by quantifying the relative contribution of ID to the total turbulent dissipation and by examining the connection between the WKE results and finescale parameterization.

\section{Methodology}
\label{sec:method}
\subsection{The wave kinetic equation}
The wave kinetic equation (WKE) describes the evolution of the wave action spectrum under interactions of weakly nonlinear waves, providing a framework for understanding the energy transfer across scales. For internal gravity waves (IGWs), the WKE is given by  
\begin{eqnarray} \label{eq:WKE}
\frac{\partial n(\mathbf{p},t)}{\partial t}
=\iint & 4\pi |V(\mathbf{p},\mathbf{p}_1,\mathbf{p}_2)|^2 \, \mathcal{F}_{p12} \, \delta(\omega-\omega_1-\omega_2)\delta(\mathbf{p}-\mathbf{p}_1-\mathbf{p}_2) \, \dd\mathbf{p}_1 \, \dd\mathbf{p}_2\\ \nonumber
-\iint & 8\pi |V(\mathbf{p}_1,\mathbf{p},\mathbf{p}_2)|^2 \, \mathcal{F}_{1p2} \, \delta(\omega-\omega_1+\omega_2)\delta(\mathbf{p}-\mathbf{p}_1+\mathbf{p}_2) \, \dd\mathbf{p}_1 \, \dd\mathbf{p}_2.
\end{eqnarray}

The right-hand side (RHS) of \eqref{eq:WKE}, namely the collision integral, describes the time evolution of the wave action density at a given wavenumber $\mathbf{p}$ due to triad interactions with two other components $\mathbf{p}_1$ and $\mathbf{p}_2$ satisfying the resonant conditions $\mathbf{p} = \mathbf{p}_1 \pm \mathbf{p}_2$ and $\omega = \omega_1 \pm \omega_2$. The functions $\mathcal{F}_{p12} = n_1 n_2 - n_p (n_1 + n_2)$ and $\mathcal{F}_{1p2} = n_p n_2 - n_1 (n_p + n_2)$ are quadratic in terms of wave action, where $n_p$ is shorthand for $n(\mathbf{p},t)$, $n_1$ for $n(\mathbf{p}_1,t)$, and so forth. The interaction kernel $V$ has been derived using various methods for hydrostatic  (e.g., \citealt{Olbers1974, Muller1975, Olbers1976, McComas1977b, Lvov2001, Lvov2004b, Lvov2010}) and non-hydrostatic  (e.g., \citealt{Labarre2024b}) setups, with the former group of results shown to be equivalent on the resonant manifold \citep{Lvov2012}.

An important metric to characterize the nonlinearity level of wave-wave interactions is the normalized Boltzmann rate \citep{Nazarenko2011, Lvov2012}, which is the ratio between the linear time scale (wave period) $\tau^\text{L}$ and the nonlinear time scale $\tau^\text{NL}$
\begin{equation} \label{eq:Bo}
Bo \equiv \frac{\tau^\text{L}}{\tau^\text{NL}}=\frac{2\pi}{\omega} \frac{\partial E/\partial t}{E}.
\end{equation}

The normalized Boltzmann rate establishes a criterion for interpreting the WKE results within specific spectral regimes. Theoretically, the WKE \eqref{eq:WKE} is valid only when $|Bo| \ll 1$, as implied by the weakly nonlinear assumption underlying wave turbulence theory \citep{Zakharov1992,Nazarenko2011}.

Although earlier results based on McComas \textit{et al.} often relied on heuristic assumptions, such as the (overly) simplified collision integral, recent advances in high-performance computing have allowed for direct evaluation of the complete collision integral for general spectral forms \citep{Eden2019a, Eden2019b, Eden2020, Dematteis2021, Dematteis2022, Dematteis2024, Wu2023b, Lanchon2023b, Labarre2024a, Labarre2024b}. The WKE has since been applied to global datasets of IGW spectra and benchmarked by finescale parameterization and microstructure observations \citep{Dematteis2024}, establishing itself as a powerful tool for estimating turbulent dissipation and improving parameterizations of ocean mixing in general circulation and climate models. In this work, we follow the numerical method in \cite{Wu2023b} for the evaluation of the collision integral, with the necessary details described in the next section.

\subsection{Numerical procedures} \label{sec:2.2}

To simulate a physical problem representative of oceanic IGWs, we consider a horizontally isotropic domain with a horizontal circular radius of 42.4 km and a vertical extent of 2.1 km. This vertical extent is chosen to minimize the effects of surface and bottom boundaries, allowing a focus on IGW interactions in the ocean interior. The wavenumber domain is discretized using a $128 \times 128$ logscale grid in both $k$ and $m$\footnote{Results using the logscale grid and the previous linear grid \citep{Wu2023b} do not show a statistically significant difference upon testing.}, with wavenumber ranges $k \in [1.5 \times 10^{-4}, 1.6 \times 10^{-1}]$ m$^{-1}$ and $m \in [3.0 \times 10^{-3}, 3.2]$ m$^{-1}$. This setup provides a spatial resolution as fine as 40 m horizontally and 2 m vertically.

The GM-like spectrum (Appendix~\ref{sec:AppA}) is used as input to the WKE \eqref{eq:WKE}, specifically in terms $\mathcal{F}_{p12}$ and $\mathcal{F}_{1p2}$. At an instantaneous time $t$, integrating the collision integral in \eqref{eq:WKE} yields $\partial n(\mathbf{p})/\partial t$ and consequently $\partial E(\mathbf{p})/\partial t$. Invoking the conservation of spectral energy
\begin{equation} \label{eq:energy}
\frac{\partial E(m)}{\partial t} + \frac{\partial \mathcal{P}(m)}{\partial m} = 0,
\end{equation}
where $\partial E(m)/\partial t =\iint (\partial E(\mathbf{p})/\partial t) \dd k_x \dd k_y$, one can define the downscale energy flux $\mathcal{P}(m)$ across an arbitrary vertical wavenumber $m$
\begin{equation} \label{eq:Pm}
\mathcal{P}(m)=\int_0^{m} \frac{\partial E(m')}{\partial t} \, \dd m' = \int_0^{m} \left[ \iint \frac{\partial E(k'_x,k'_y,m')}{\partial t} \, \dd k'_x \, \dd k'_y \right] \, \dd m'.
\end{equation} 


Instead of directly resolving turbulent events, the WKE evaluates the energy flux down to the 10-meter vertical scale (represented by the critical vertical wavenumber $m_\text{c} = 0.62$ m$^{-1}$) as an estimate of the energy available for turbulent dissipation. However, interpreting the WKE results near $m_\text{c}$ is often constrained by potential violation of the weakly nonlinear assumption \citep{Holloway1978, Holloway1980}. 
Although recent studies have shown that $\mathcal{P}(m)$ exhibits low sensitivity to $m$ near $m_\text{c}$ \citep{Wu2023b, Dematteis2024}, we choose to further quantify this uncertainty by introducing a spectrum-specific cutoff vertical wavenumber $m_\text{cutoff}$ (usually less than $m_\text{c}$), up to which no more than 10\% of waves violate the weakly nonlinear assumption, characterized by $|Bo| > 0.2$. 
It is important to acknowledge the gap between $m_\text{cutoff}$, beyond which the WKE becomes invalid, and $m_\text{c}$, at which IGWs become unstable to shear instability. Turbulent dissipation is approximated as the mean value of the downscale energy flux $\mathcal{P}(m)$ over the range $m \in [m_\text{cutoff}$, $m_\text{c}]$ if $m_\text{cutoff} < m_\text{c}$. The difference between the maximum and minimum values in $\mathcal{P}(m)$ over this range is introduced as the uncertainty. 
When $m_\text{cutoff} > m_\text{c}$, the uncertainty is zero. 

To evaluate the relative contribution of ID to the total turbulent dissipation, we isolate ID triads by applying a selection criterion based on the geometry of individual triads. We rank the frequencies of each wave component in a triad from high to low as $(\omega^H,\omega^M,\omega^L)$ and the magnitudes of vertical wavenumbers as $(|m^H|,|m^M|,|m^L|)$.
As a scale-separated mechanism in both $\omega$ and $m$, an ID triad consists of a low-$\omega$, low-$m$ wave and two high-$\omega$, high-$m$ waves (FIG.~\ref{fig:2}), satisfying $\omega^M/\omega^L > 2$ and $|m^M|/|m^L| > 2$. The threshold value for ``scale separation'' is set at 2, as in \citet{Wu2023b}. 
Similar selection procedures have been adopted by \citet{Eden2019b} and \citet{Dematteis2024}.

\section{Results}
\label{sec:results}

We start with the GM spectrum \citep{Cairns1976} and then extend to spectra that deviate from GM, with a focus on the role of ID in turbulent dissipation across varying spectra. The direction of action diffusion, as described by \eqref{eq:ID}, depends on the sign of $\sigma$, which is the vertical-wavenumber spectral slope of the action spectrum, $n \propto m^\sigma$. For GM-like spectra, $\sigma \equiv s_m - s_\omega$ represents the difference between the vertical-wavenumber and frequency spectral slopes of the energy spectrum (see Appendix~\ref{sec:AppA} for a detailed illustration). We consider the range $\sigma \in [-0.5, 0.5]$, corresponding to $s_m \in [-2.5, -1.5]$ with fixed $s_\omega = -2$, which is consistent with the range from global statistics of field measurements \citep{Dematteis2024}. The energy level, $E_0 = 3\times 10^{-3}$ m$^{-2}$ s$^{-2}$, as defined in \eqref{eq:GM}, is kept constant as $\sigma$ varies to ensure that the total energy of the IGW field remains unchanged. For comparison with GM, we present two extreme cases: a red spectrum with $\sigma=-0.5$ and a blue spectrum with $\sigma=0.5$. We then explore the entire range of $\sigma \in [-0.5, 0.5]$, followed by a sensitivity study with respect to the parameters $E_0$ and $s_\omega$ in Appendix~\ref{sec:AppB}.

In the absence of generation and dissipation of IGWs, energy is conserved within finite domain and redistributes through wave-wave interaction, with energy fluxes across the (spectral) domain boundaries remaining zero. Therefore, the rate of change of spectral energy density $\partial E/\partial t$ interprets spectral energy transfer within the domain, as shown in \eqref{eq:energy}. 
Henceforth, we define regimes where $\partial E/\partial t < 0$ as sources, since their energy decays over time, supplying energy to other regimes. Conversely, regimes where $\partial E/\partial t > 0$ are sinks, since their energy increases, accumulating energy. The terms ``source'' and ``sink'' follow the conventions of \citet{Eden2019a, Eden2019b}, where they describe the direction of energy transfer rather than referring to specific generation or dissipation mechanisms. 

\subsection{The standard GM spectrum} \label{sec:3.1}

\begin{figure}
\centering
\includegraphics[width=\textwidth]{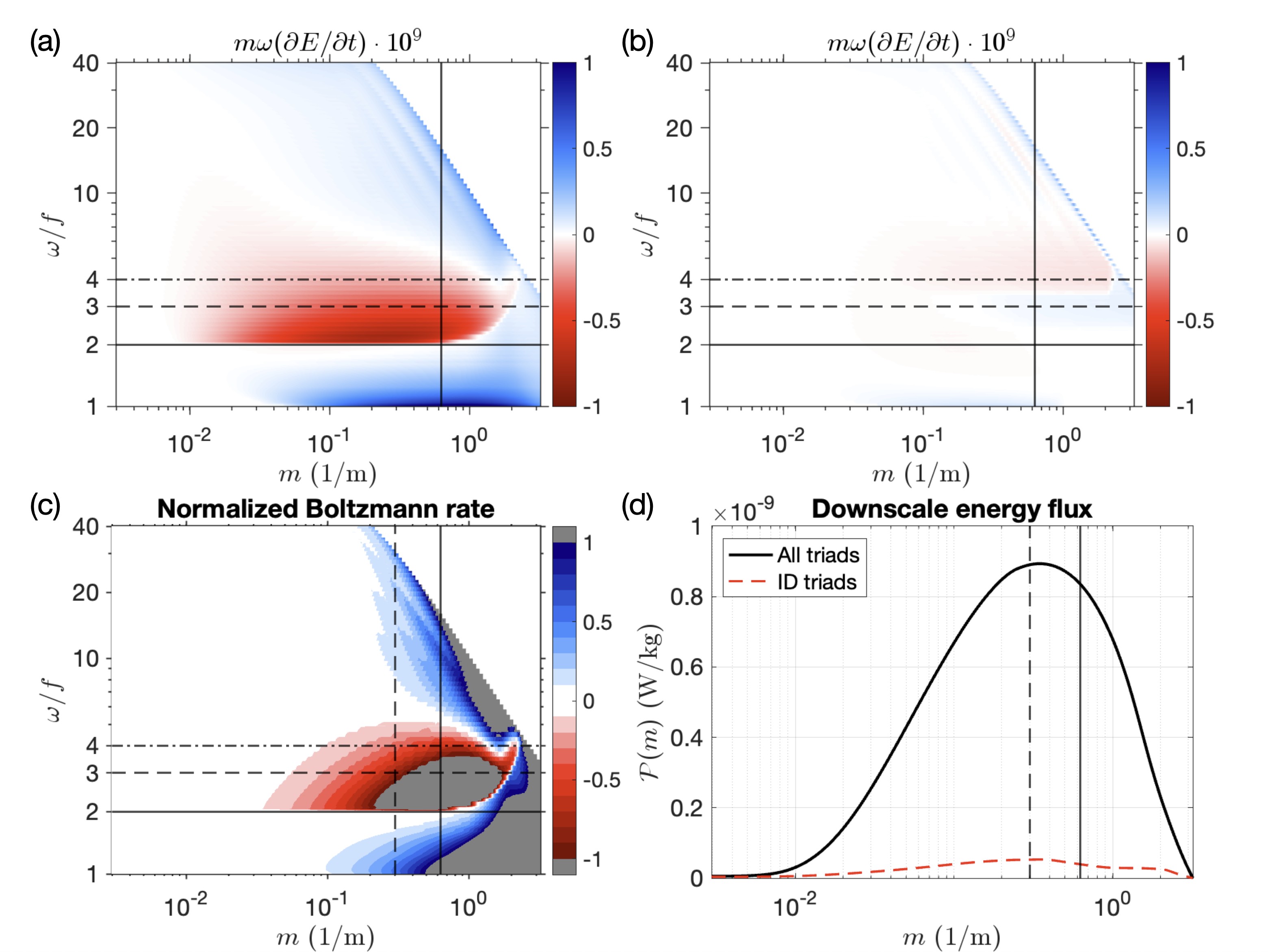}
\caption{(\textit{a}) Spectral energy transfer, $m\omega \,(\partial E/\partial t)$, computed from the WKE \eqref{eq:WKE} for the Garrett-Munk spectrum of the form $n \propto  m^0$ in the high-$m$, high-$\omega$ limit. The prefactor $m\omega$ is included to preserve variance in the log-log representation. Energy sources ($\partial E/\partial t < 0$) and sinks ($\partial E/\partial t > 0$) are indicated in red and blue, respectively. 
(\textit{b}) Same as (\textit{a}), but retaining only the induced diffusion (ID) mechanism. 
(\textit{c}) Normalized Boltzmann rate \eqref{eq:Bo}, where $|Bo| \ll 1$ indicates weak nonlinearity and the validity of the WKE. 
(\textit{d}) Downscale energy flux \eqref{eq:Pm}, shown for all triads and for ID triads only.
Horizontal lines in \textit{(a--c)} denote frequencies $2f$, $3f$, and $4f$. Vertical solid and dashed lines denote the critical vertical wavenumber $m_\text{c}$ and the cutoff vertical wavenumber $m_\text{cutoff}$, respectively.}
\label{fig:3}
\end{figure}

For the GM spectrum characterized by a white action spectrum $n \propto m^0$, spectral energy transfer arising from all triad interactions exhibits a source between $2f$ and $4f$, with sinks at lower and higher frequencies [FIG.~\ref{fig:3}(\textit{a})]. The normalized Boltzmann rate \eqref{eq:Bo} indicates that the high-$m$ regime is subject to strong nonlinearity, casting doubt on the validity of the WKE in this regime [FIG.~\ref{fig:3}(\textit{c})]. For GM, the corresponding cutoff vertical wavenumber is $m_\text{cutoff} = 0.30$ m$^{-1}$, a factor of 2 smaller than the critical vertical wavenumber $m_\text{c} = 0.62$ m$^{-1}$. 
The WKE estimates the total turbulent dissipation $\mathcal{P}_\text{WKE} = (8.12 \pm 0.26) \times 10^{-10}$ W kg$^{-1}$ [FIG.~\ref{fig:3}(\textit{d})]\footnote{These results correspond to a Coriolis frequency $f=7.84\times 10^{-5}$ s$^{-1}$ and a buoyancy frequency $N=5.24\times 10^{-3}$ s$^{-1}$ (Appendix~\ref{sec:AppA}), which differ from the values used in \cite{Wu2023b}.}, in good agreement with the finescale parameterization prediction $\mathcal{P}_\text{FP}=8 \times 10^{-10}$ W kg$^{-1}$, with the latter computed following the standard procedure described in \cite{Polzin2014}.
Due to the vanishing gradient of the action spectrum in $m$, ID contributes almost no flux, except for some weak secondary diffusion [FIG.~\ref{fig:3}(\textit{b,d})], corroborating the findings of \citet{Dematteis2022}.

\subsection{A red action spectrum} \label{sec:3.2.1}

\begin{figure}
\centering
\includegraphics[width=\textwidth]{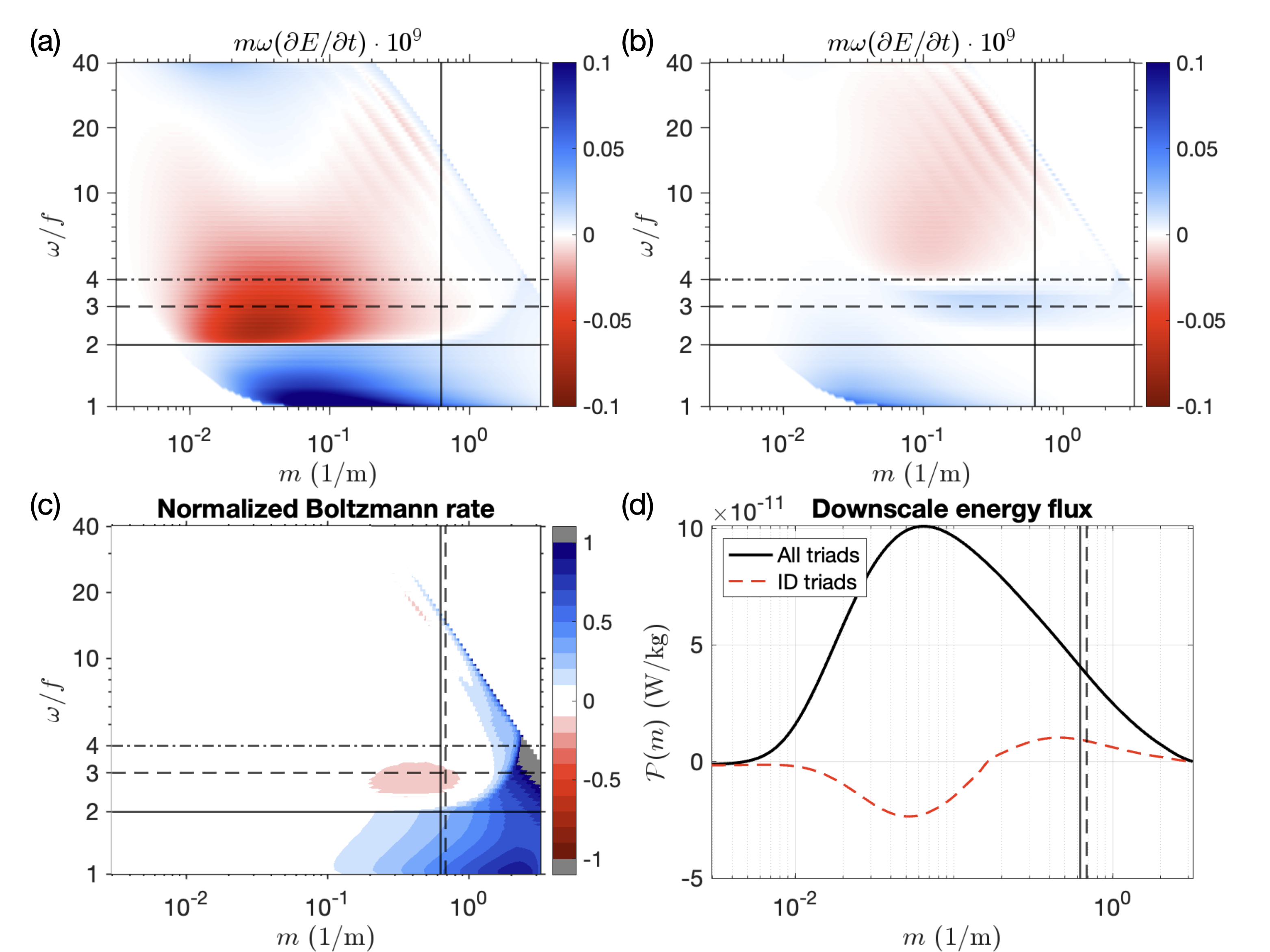}
\caption{Same as FIG.~\ref{fig:3} but for a red action spectrum, $n \propto m^{-0.5}$ in the high-$m$, high-$\omega$ limit.}
\label{fig:4}
\end{figure}

For a typical red action spectrum characterized by $n \propto m^{-0.5}$, action and energy are more concentrated at large vertical scales compared to GM. With total energy held constant, action and energy at small scales are correspondingly reduced. Spectral energy transfer is dominated by a source between $2f$ and approximately $10f$, with a sink below $2f$ and a much weaker sink above $20f$ [FIG.~\ref{fig:4}(\textit{a})]. The magnitudes of the source and sinks, along with the downscale energy flux, are an order of magnitude smaller than those in GM [FIG.~\ref{fig:4}(\textit{a,d})]. The weakly nonlinear assumption is better satisfied compared to that for GM, as indicated by the normalized Boltzmann rate \eqref{eq:Bo} [FIG.~\ref{fig:4}(\textit{c})]. The corresponding cutoff vertical wavenumber is $m_\text{cutoff} = 0.69$ m$^{-1}$, allowing the WKE results to extend to the dissipation scale represented by $m_\text{c} = 0.62$ m$^{-1}$ without introducing uncertainties as described in \S\ref{sec:2.2}.

Induced diffusion exhibits a source above $4f$ and two distinct sinks below $4f$ [FIG.~\ref{fig:4}(\textit{b})]. The two sinks occur in separate regimes relative to the source: the first sink spreads over intermediate frequencies ($\omega \approx 3f$) and large vertical wavenumbers up to $m \sim \textit{O}(1)$ m$^{-1}$, while the second sink is concentrated near the inertial frequency ($\omega \approx f$) and small vertical wavenumbers ($m \lesssim 0.1$ m$^{-1}$). These three regimes---comprising the source and the two sinks---reflect the diffusive and scale-separated transfers characteristic of ID. In particular, the source and the first sink arise from the action diffusion at small scales described by \eqref{eq:ID}, featuring a forward cascade in $m$ accompanied by a backward cascade in $\omega$. Since this diffusive transfer conserves action, it results in an energy surplus when moving towards lower frequencies, as energy is given by $E=n\omega$, and $\omega$ decreases\footnote{This is more straightforward if we focus on a single triad (e.g., the one in FIG.~\ref{fig:2}). In this case, a red action spectrum leads to a forward diffusive transfer from $\mathbf{p}$ to $\mathbf{p}_2$. While action is conserved, i.e., $\Delta n =-\Delta n_2$, energy is not: the energy lost by $\mathbf{p}$ is always greater than that received by $\mathbf{p}_2$, i.e., $\omega\Delta n > \omega_2(-\Delta n_2)$, since $\omega > \omega_2$. This results in an energy surplus between $\mathbf{p}$ and $\mathbf{p}_2$, where excess energy shall be absorbed by the largescale mode $\mathbf{p}_1$, indicative of a backward scale-separated transfer.}. To conserve total energy, excess energy must be absorbed by the large scale, leading to the formation of the second sink and enabling a scale-separated transfer which is backward in both $m$ and $\omega$.

Inclusion of both diffusive and scale-separated transfers is crucial to understanding the full picture of ID, as further illustrated by the ID-driven downscale energy flux $\mathcal{P}^{\text{ID}}(m)$ [FIG.~\ref{fig:4}(\textit{d})]. ID represents a backward cascade with $\mathcal{P}^{\text{ID}}(m) < 0$ when $m < 0.15$ m$^{-1}$ and a forward cascade with $\mathcal{P}^{\text{ID}}(m) > 0$ when $m > 0.15$ m$^{-1}$. The former results from the scale-separated transfer with the large scale as a sink, and the latter is governed by the diffusive transfer described by \eqref{eq:ID}. At dissipation scale, the downscale energy flux is dominated by the diffusive transfer with $\mathcal{P}^{\text{ID}}(m_\text{c}) = (0.06 \pm 0.00) \times 10^{-10}$ W kg$^{-1}$, which contributes to mixing.

\subsection{A blue action spectrum} \label{sec:3.2.2}

\begin{figure}
\centering
\includegraphics[width=\textwidth]{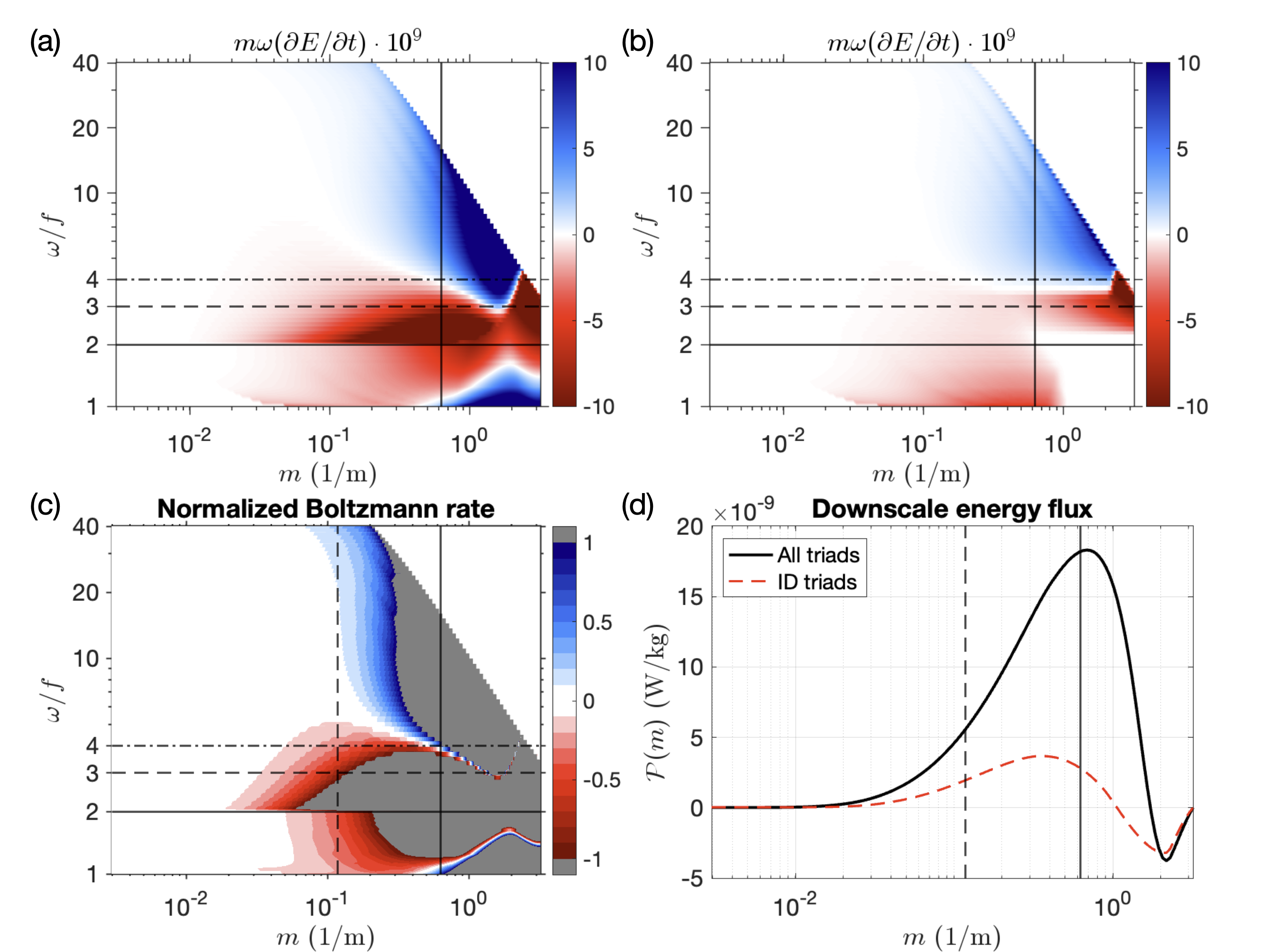}
\caption{Same as FIG.~\ref{fig:3} but for a blue action spectrum, $n \propto m^{0.5}$ in the high-$m$, high-$\omega$ limit.}
\label{fig:5}
\end{figure}

For a typical blue action spectrum characterized by $n \propto m^{0.5}$, more action and energy are distributed to small vertical scales compared to GM. Spectral energy transfer is dominated by a source below $4f$, with sinks below $1.5f$ and above $3f$, concentrated at $m > \textit{O}(1)$ m$^{-1}$ [FIG.~\ref{fig:5}(\textit{a})]. The magnitudes of the source and sinks, along with the downscale energy flux, are an order of magnitude greater than those in GM [FIG.~\ref{fig:5}(\textit{a,d})]. Violation of the weakly nonlinear assumption is more pronounced, as indicated by the normalized Boltzmann rate \eqref{eq:Bo} [FIG.~\ref{fig:5}(\textit{c})]. The corresponding cutoff vertical wavenumber $m_\text{cutoff} \approx 0.12$ m$^{-1}$ is significantly smaller than the critical vertical wavenumber $m_\text{c} \approx 0.62$ m$^{-1}$, resulting in increased uncertainty in turbulent dissipation estimates [FIG.~\ref{fig:5}(\textit{d})]. 

Induced diffusion manifests the reversed scenario relative to the red spectral case presented in \S\ref{sec:3.2.1}. In particular, the diffusive transfer at small scales is now backward toward lower $m$, and the scale-separated transfer is forward with energy sourced from the large scale to compensate for the deficit at small scales\footnote{In this case, a blue action spectrum leads to a backward diffusive transfer from mode $\mathbf{p}_2$ to mode $\mathbf{p}$ in a single triad (FIG.~\ref{fig:2}). The energy required by $\mathbf{p}$ is always greater than that supplied by $\mathbf{p}_2$, i.e., $\omega\Delta n > \omega_2(-\Delta n_2)$, given that $\omega > \omega_2$ and $\Delta n =-\Delta n_2$. This results in an energy deficit between $\mathbf{p}$ and $\mathbf{p}_2$, which shall be compensated for by the largescale mode $\mathbf{p}_1$, indicative of a forward scale-separated transfer.} [FIG.~\ref{fig:5}(\textit{b})]. However, the regime where the backward cascade dominates is confined to vertical scales smaller than the dissipation scale, while the forward cascade regime spans a much broader range of $m$, encompassing both $m_\text{cutoff}$ and $m_\text{c}$. 

The downscale energy flux driven by ID is quantified as $\mathcal{P}^\text{ID} = (20.99 \pm 3.10) \times 10^{-10}$ W kg$^{-1}$ over the range between $m_\text{cutoff}$ and $m_\text{c}$ [FIG.~\ref{fig:5}(\textit{d})]. Within this range, the forward scale-separated transfer dominates the ID cascade, with energy fluxed downscale to sustain turbulent dissipation. As a result, despite its reversed direction relative to the red spectral case, ID continues to act as a dissipative mechanism.

\section{Discussion and conclusion}
\label{sec:conclusion}

\begin{figure}
\centering
\includegraphics[width=.8\textwidth]{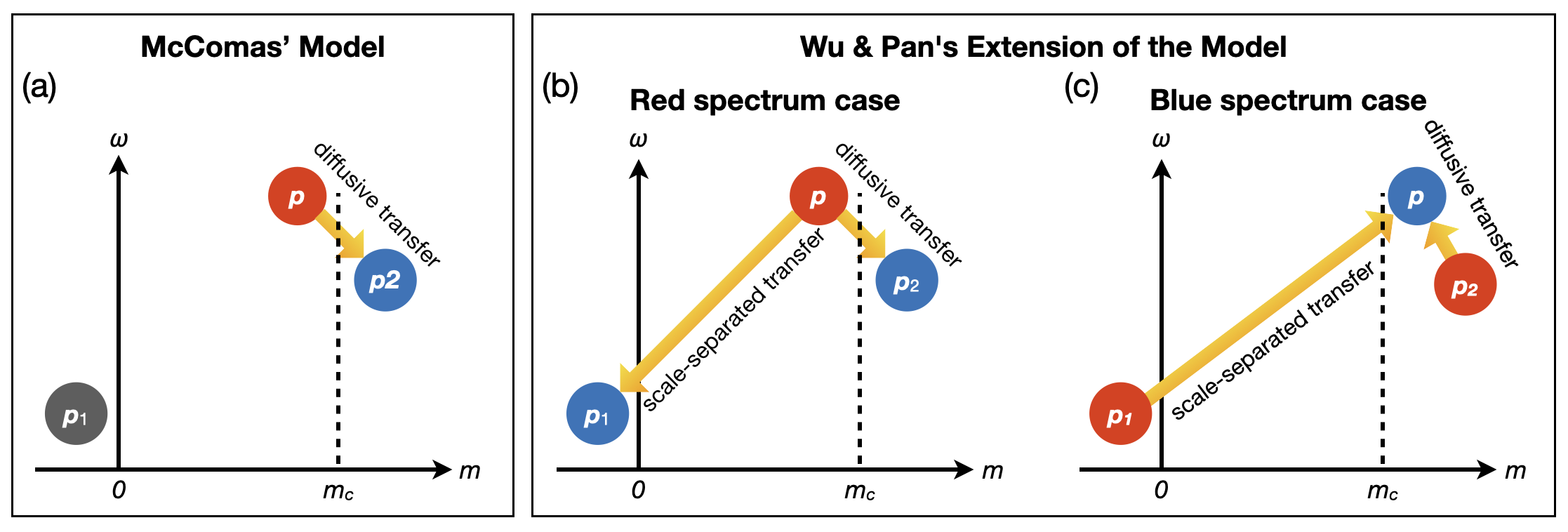}
\caption{Conceptual models of induced diffusion (ID) for a resonant wave triad $\mathbf{p} = \mathbf{p}_1 + \mathbf{p}_2$. The largescale, near-inertial wave $\mathbf{p}_1$ has an oppositely signed vertical wavenumber and thus appears to the left of the $m = 0$ axis (FIG.~\ref{fig:2}). 
(\textit{a}) McComas’ model illustrating a diffusive transfer from $\mathbf{p}$ to $\mathbf{p}_2$ at small scales, while the largescale wave $\mathbf{p}_1$ remains stationary. 
(\textit{b--c}) Our extension of the model for red ($\sigma < 0$) and blue ($\sigma > 0$) action spectra, $n \propto m^\sigma$, respectively. In both cases, a diffusive transfer (between $\mathbf{p}$ and $\mathbf{p}_2$) as well as a scale-separated transfer (involving $\mathbf{p}_1$) are highlighted, presenting ID as a broadband process rather than one confined to small scales. 
Red, blue, and gray dots denote energy sources, sinks, and stationary states, respectively. Yellow arrows indicate the direction of energy transfer. 
Turbulent dissipation is approximated by the downscale energy flux across the critical vertical wavenumber $m_\text{c}$.}
\label{fig:6}
\end{figure}

We begin by elaborating on and summarizing the role of induced diffusion (ID) in ocean mixing for spectra deviating from the Garrett-Munk (GM) spectrum. At a given vertical wavenumber, such as $m_\text{c}$, the energy flux comprises two components: a diffusive transfer described by \eqref{eq:ID} and a scale-separated transfer associated with energy absorption or compensation by the large scale. A complete picture of ID in the spatiotemporal domain is illustrated in FIG.~\ref{fig:6}. For red action spectra with $\sigma<0$ in $n \propto m^\sigma$, the diffusive transfer dominates near $m_\text{c}$, driving a forward cascade toward dissipation. As $\sigma$ increases and crosses zero (corresponding to blue spectra), the scale-separated transfer becomes more and more important near $m_\text{c}$, where the energy compensation process supplies energy available for dissipation. While McComas's original conceptualization of ID was based on a frozen largescale field [FIG.~\ref{fig:6}(\textit{a})], our evaluation of the wave kinetic equation (WKE) reveals far richer dynamics of ID, in which the large scale actively participates in the energy cascade and plays a crucial role in driving turbulent dissipation [FIG.~\ref{fig:6}(\textit{b,c})].


\begin{figure}
\centering
\includegraphics[width=1\textwidth]{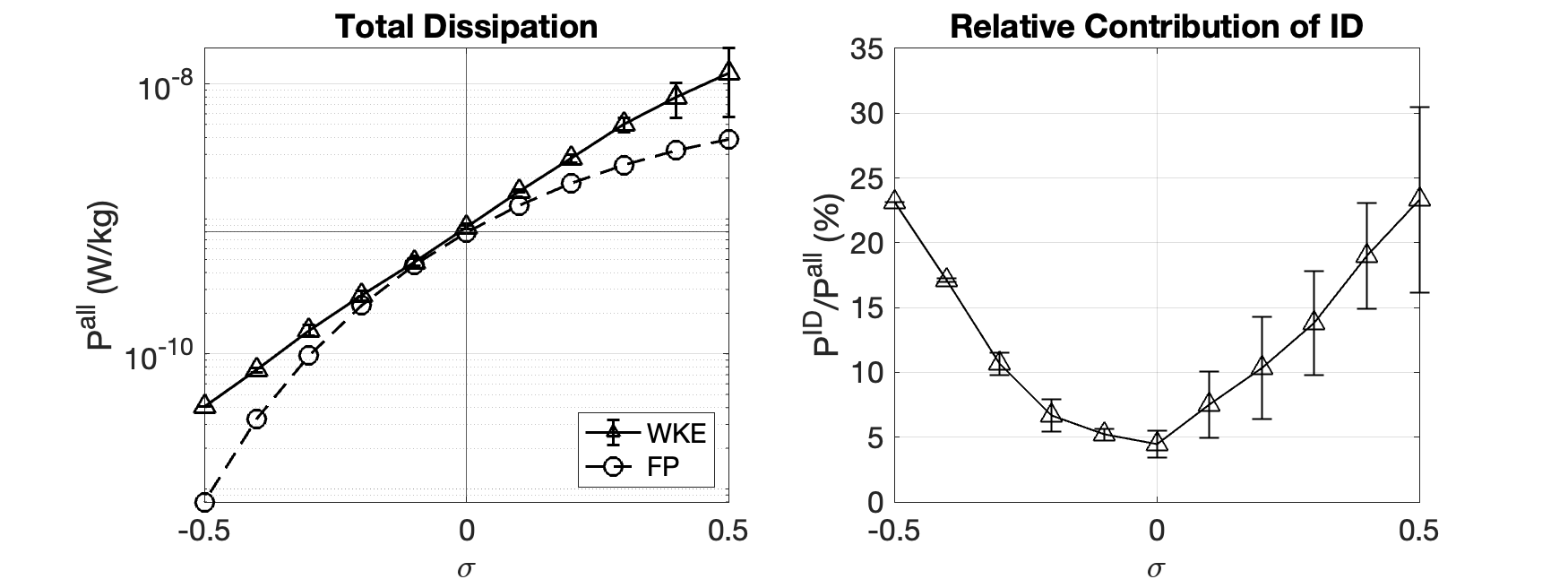}
\caption{(Left) Total turbulent dissipation $\mathcal{P}^\text{all}$ estimated using the wave kinetic equation (WKE) compared with that obtained from finescale parameterization (FP). 
(Right) Relative contribution of induced diffusion (ID), $\mathcal{P}^\text{ID}/\mathcal{P}^\text{all}$, as a function of $\sigma \equiv s_m - s_\omega$. All results are based on a fixed energy level of $E_0 = 3 \times 10^{-3}$ m$^{-2}$ s$^{-2}$ and a constant frequency spectral slope of $s_\omega = -2.0$.
The error bars represent the uncertainty in $\mathcal{P}^\text{all}(m)$ and $\mathcal{P}^\text{ID}(m)/\mathcal{P}^\text{all}(m)$ over the range $m \in [m_\text{cutoff}$, $m_\text{c}]$ if $m_\text{cutoff} < m_\text{c}$. When $m_\text{cutoff} > m_\text{c}$, the uncertainty is zero. }
\label{fig:7}
\end{figure}

For a comprehensive evaluation of the role of ID across varying spectra, we swept the entire range of $\sigma \in [-0.5, 0.5]$ using a step size of 0.1. The total turbulent dissipation increases dramatically with increasing $\sigma$, as bluer spectra allocate more energy to small scales and thus drive stronger turbulent dissipation (FIG.~\ref{fig:7}).
The relative contribution of ID is quantified as the turbulent dissipation driven by ID triads normalized by the total turbulent dissipation by all triads. For $\sigma = 0$, i.e., the GM spectrum, $\mathcal{P}^\text{ID}/\mathcal{P}^\text{all}$ is minimal due to the vanishing of both diffusive and scale-separated transfers; in this case, ID contributes almost no flux, despite some weak secondary diffusion \citep{Dematteis2022, Wu2023b}. Apart from this state, the relative contribution of ID remains consistently positive, and the ratio $\mathcal{P}^\text{ID}/\mathcal{P}^\text{all}$ is positively correlated with the deviation $|\sigma|$. At the two endpoints, $\sigma = \pm 0.5$, ID contributes up to 25\% of the total dissipation. Moreover, the associated uncertainty grows with increasing $\sigma$, reflecting the widening gap between $m_\text{cutoff}$ and $m_\text{c}$. 

This study represents a significant step forward in understanding the role of ID in oceanic mixing. Unlike the empirical approach of finescale parameterization, the WKE captures the underlying mechanisms of wave-wave interactions, enabling diagnostic insights such as the role of ID, which constitutes the central focus of this study. Using the WKE, we address long-standing theoretical gaps and provide a physically grounded depiction of ID in the spatiotemporal domain, without being restricted to the high-$\omega$, high-$m$ regime of the spectra or relying on the diffusion equation \eqref{eq:ID} as a reduced-order alternative for the WKE \eqref{eq:WKE}. By elucidating the dynamics of energy cascade in the IGW field, our findings offer valuable insights into the specific types of wave-wave interactions responsible for turbulent dissipation.


We conclude by placing two caveats on the present work. First, the analysis is based on the instantaneous energy transfer of GM-like spectra. The underlying assumption is that these spectra remain stationary under balanced forcing and dissipation in the ocean. 
An important direction is to examine the evolution of the spectra under wave-wave interactions within the WKE framework (see the recent work by \citealt{Labarre2025}). Second, the present study assumes that the spectra retain a power-law form at small scales, even beyond the dissipation scale $m_\text{c}$. 
In practice, this assumption may be violated due to dissipative effects, which could result in a damped spectral tail beyond $m_\text{c}$ and thus affect the interpretation of ID, especially in the case of blue spectra. 
A detailed investigation of this issue likely requires simulations of stratified turbulence. We leave this opportunity to future research.

\backsection[Funding]{This research is supported by the National Science Foundation (award number OCE-2241495 and OCE-2306124) and the Simons Foundation through Simons Collaboration on Wave Turbulence.}

\backsection[Declaration of interests]{The authors report no conflict of interest.}

\backsection[Data availability statement]{The data that support the findings of this study are openly available on GitHub at https://github.com/yue-cynthia-wu.}




\appendix
\section{The Garrett-Munk (GM) spectrum and variations} \label{sec:AppA}

The spectral representation of oceanic IGWs was first modeled by Peter H. Garrett and Walter H. Munk in the 1970s in a series of publications \citep{Garrett1972, Garrett1975, Cairns1976}, providing a statistical and empirical description of the wave energy distribution based on the frequency and vertical wavenumber of IGWs
\begin{equation} \label{eq:GM} 
E(\omega,m)=\frac{N}{N_0}\,E_0 \,A(m)\,B(\omega),
\end{equation}
where $E(\omega,m)$ is the wave energy in the frequency-vertical wavenumber domain. The factor $N/N_0$ is a stratification scaling, where $N$ and $N_0=5.24\times 10^{-3}$ s$^{-1}$ are the actual and reference buoyancy frequencies, respectively. The parameter $E_0$ is the energy level of the IGW field. 

Functions $A$ and $B$ in \eqref{eq:GM} are separable with respect to $m$ and $\omega$ and are normalized to integrate to unity such that the total energy $\iint E(\omega,m) \,\dd\omega\,\dd m=(N/N_0)E_0$
\begin{equation} \label{eq:A} 
A(m) \propto \frac{1}{m^*}\left[1+\left(\frac{m}{m^*}\right)^r \right]^{s_m/r},
\end{equation}
\begin{equation} \label{eq:B}
B(\omega) \propto \omega \, ^{s_\omega-2s_\text{NI}}\left(\omega^2-f^2 \right)^{s_\text{NI}},
\end{equation}
where $m^* = \pi j/b$ is the characteristic vertical wavenumber, $j$ is the mode number, and $b$ is the stratification scale height. 
The parameter $r$ controls the steepness of the transition from the low-$m$ plateau to the high-$m$ power-law regime; $r = 2$ is commonly used, as alternative values are rarely confirmed observationally. 
The Coriolis or inertial frequency $f = 2 \Omega \sin \varphi$ is a function of latitude $\varphi$, where $\Omega = 7.29 \times 10^{-5}$ s$^{-1}$ is the Earth's rotational angular velocity. 
The spectrum is characterized by three spectral slopes: $s_\text{NI}$ in the near-inertial frequency limit, $s_\omega$ in the high-$\omega$ limit, and $s_m$ in the high-vertical-wavenumber limit.
Common to all variations of GM is the presence of an inertial peak and red spectra in both $\omega$ and $m$, signifying a concentration of energy near the inertial frequency and in low vertical modes \citep{Polzin2011}.

For the standard GM spectrum described in \citet{Cairns1976}, the parameters $j=4$ and $b=1300$ m so that $m^*=0.01$ m$^{-1}$. The buoyancy frequency $N=N_0=5.24\times 10^{-3}$ s$^{-1}$, and the Coriolis frequency $f=7.84\times 10^{-5}$ s$^{-1}$ corresponding to the latitude $\varphi=32.5^{\circ}$ for mid-latitude oceans. The energy level $E_0=3\times 10^{-3}$ m$^{-2}$ s$^{-2}$. The three spectral slopes are $s_\omega=s_m=-2$ and $s_\text{NI}=-0.5$.

The energy spectrum given by \eqref{eq:GM} follows power-law scaling in the high-$\omega$, high-$m$ regime, expressed as $E(\omega,m)\propto \omega^{s_\omega} m^{s_m}$. This corresponds to an action spectrum $n(k,m) \propto k^{s_\omega-2} m^{s_m-s_\omega}$. The conversion adheres to the relationship 
\begin{equation} \label{eq:conv}
n(k,m) = \frac{E(k,m)}{\omega}=\frac{E(\omega,m)}{2\pi k \omega} \frac{\partial \omega}{\partial k} = \frac{E(\omega,m)}{2\pi k \omega} \frac{N^2-\omega^2}{k^2+m^2}.
\end{equation}

Note that the relation $n=E/\omega$ holds strictly in the $(k,m)$-domain and does not apply in the $(\omega,m)$-domain. 


\section{Sensitivity study to parameters $E_0$ and $s_\omega$} \label{sec:AppB}

\begin{figure}
\centering
\includegraphics[width=.5\textwidth]{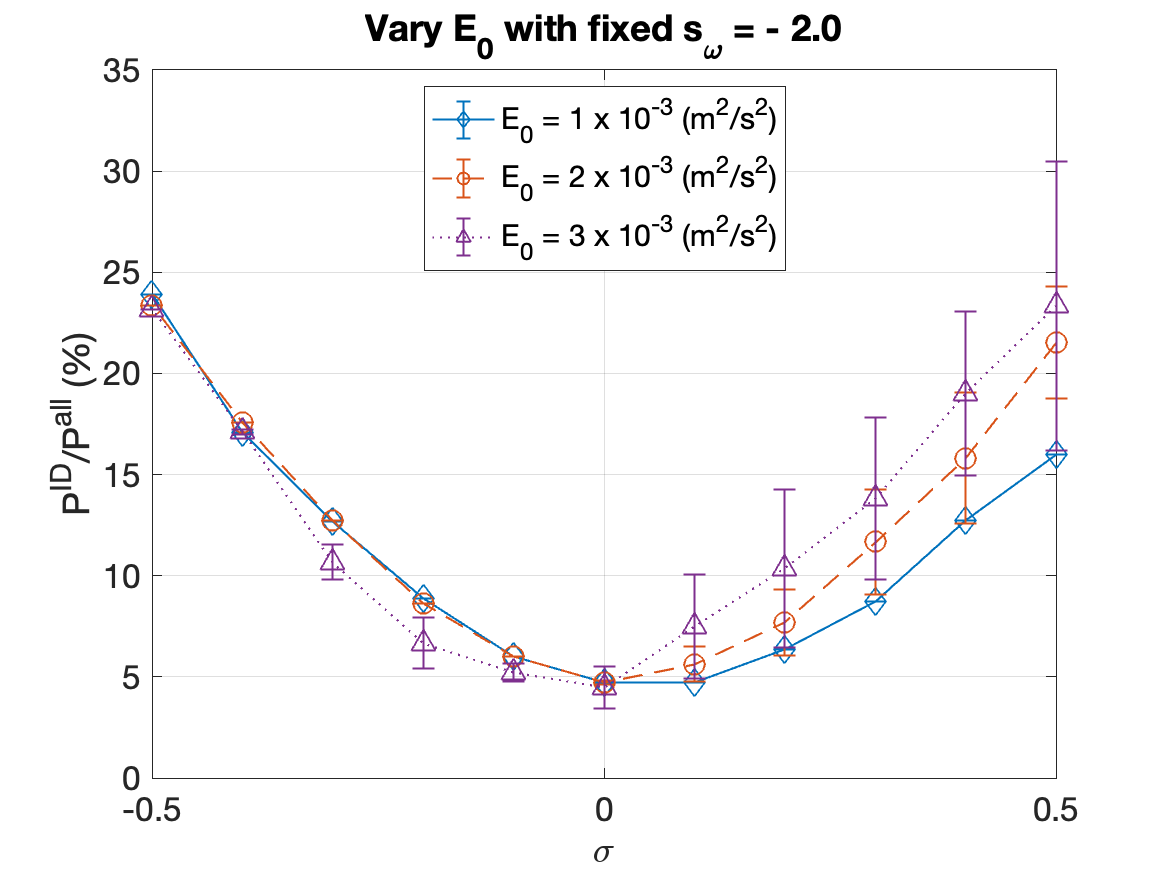}\includegraphics[width=.5\textwidth]{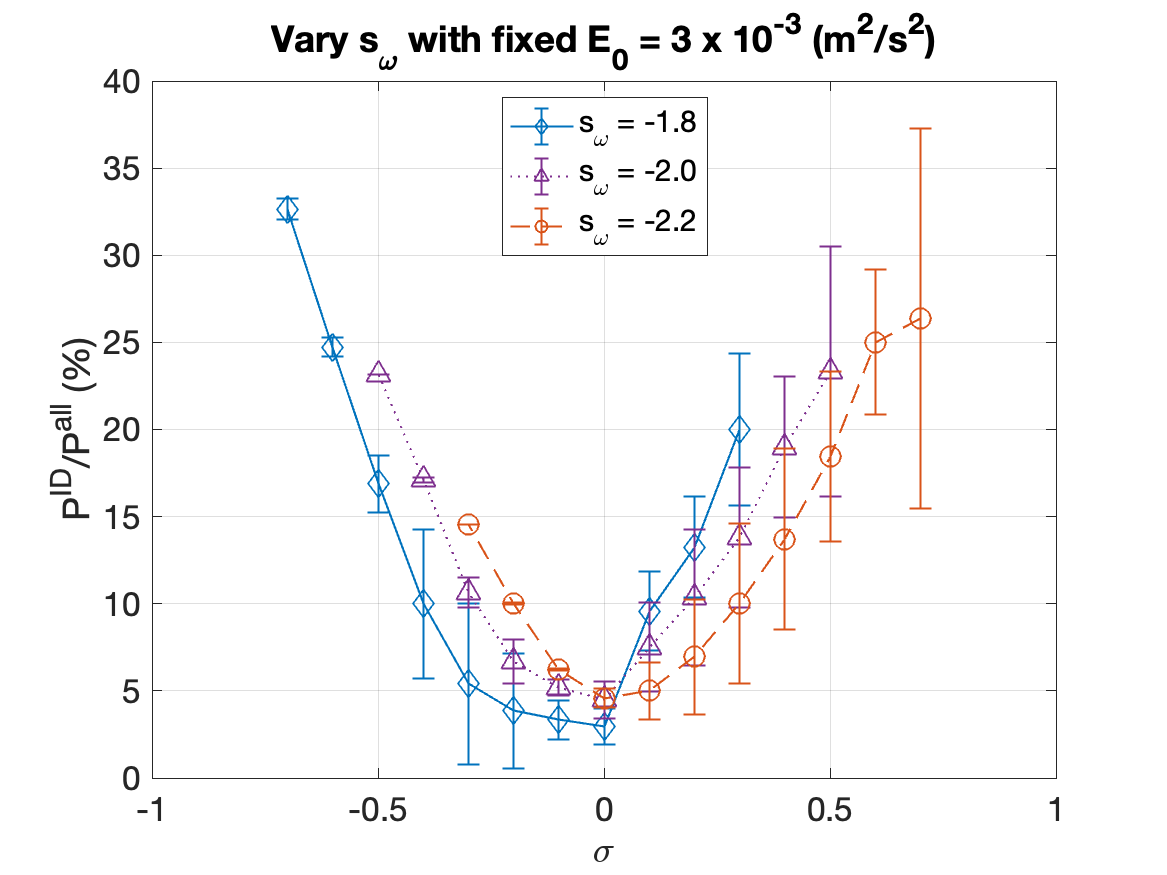}
\caption{Relative contribution of induced diffusion (ID), $\mathcal{P}^\text{ID}/\mathcal{P}^\text{all}$, as a function of $\sigma \equiv s_m - s_\omega$. (Left) Results for varying energy level $E_0$ with fixed frequency spectral slope $s_\omega = -2.0$. (Right) Results for varying $s_\omega$ with fixed $E_0 = 3 \times 10^{-3}$ m$^{-2}$ s$^{-2}$.}
\label{fig:8}
\end{figure}

The results presented in \S\ref{sec:results} are based on fixed values for the energy level $E_0=3\times 10^{-3}$ m$^{-2}$ s$^{-2}$ and the frequency spectral slope $s_\omega=-2.0$. To validate our conclusions across a broader parameter space, a sensitivity study has been conducted (FIG.~\ref{fig:8}). Considering both diffusive and scale-separated transfers, induced diffusion always contributes positively to turbulent dissipation, acting as a dissipative mechanism.

\bibliographystyle{jfm}
\bibliography{library}

\begin{thebibliography}{45}
\expandafter\ifx\csname natexlab\endcsname\relax\def\natexlab#1{#1}\fi
\def\au#1{#1} \def\ed#1{#1} \def\yr#1{#1}\def\at#1{#1}\def\jt#1{\textit{#1}} \def\bt#1{#1}\def\bvol#1{\textbf{#1}} \def\vol#1{#1} \def\pg#1{#1} \def\publ#1{#1}\def\arxiv#1{#1}\def\org#1{#1}\def\st#1{\textit{#1}}

\bibitem[Cairns \& Williams(1976)]{Cairns1976}
{\sc \au{Cairns, James~L} \& \au{Williams, Gordon~O}} \yr{1976}  \at{{Internal wave observations from a midwater float, 2}}.  \jt{J. Geophys. Res.}  \bvol{81}~(12),  \pg{1943--1950}.

\bibitem[Delpech {\em et~al.\/}(2024)Delpech, Barkan, Srinivasan, McWilliams, Arbic, Siyanbola \& Buijsman]{Delpech2024}
{\sc \au{Delpech, Audrey}, \au{Barkan, Roy}, \au{Srinivasan, Kaushik}, \au{McWilliams, James~C.}, \au{Arbic, Brian~K.}, \au{Siyanbola, Oladeji~Q.} \& \au{Buijsman, Maarten~C.}} \yr{2024}  \at{Eddy--internal wave interactions and their contribution to cross-scale energy fluxes: A case study in the california current}.  \jt{Journal of Physical Oceanography}  \bvol{54}~(3),  \pg{741 -- 754}.

\bibitem[Dematteis {\em et~al.\/}(2024)Dematteis, Le~Boyer, Pollmann, Polzin, Alford, Whalen \& Lvov]{Dematteis2024}
{\sc \au{Dematteis, Giovanni}, \au{Le~Boyer, Arnaud}, \au{Pollmann, Friederike}, \au{Polzin, Kurt~L.}, \au{Alford, Matthew~H.}, \au{Whalen, Caitlin~B.} \& \au{Lvov, Yuri~V.}} \yr{2024}  \at{Interacting internal waves explain global patterns of interior ocean mixing}.  \jt{Nature Communications}  \bvol{15}~(1),  \pg{7468}.

\bibitem[Dematteis \& Lvov(2021)]{Dematteis2021}
{\sc \au{Dematteis, Giovanni} \& \au{Lvov, Yuri~V.}} \yr{2021}  \at{Downscale energy fluxes in scale-invariant oceanic internal wave turbulence}.  \jt{Journal of Fluid Mechanics}  \bvol{915},  \pg{A129}.

\bibitem[Dematteis {\em et~al.\/}(2022)Dematteis, Polzin \& Lvov]{Dematteis2022}
{\sc \au{Dematteis, Giovanni}, \au{Polzin, Kurt} \& \au{Lvov, Yuri~V.}} \yr{2022}  \at{On the origins of the oceanic ultraviolet catastrophe}.  \jt{Journal of Physical Oceanography}  \bvol{52}~(4),  \pg{597--616}.

\bibitem[Dong {\em et~al.\/}(2020)Dong, B{\"u}hler \& Smith]{Dong2020}
{\sc \au{Dong, Wenjing}, \au{B{\"u}hler, Oliver} \& \au{Smith, K~Shafer}} \yr{2020}  \at{Frequency diffusion of waves by unsteady flows}.  \jt{Journal of Fluid Mechanics}  \bvol{905},  \pg{R3}.

\bibitem[Dong {\em et~al.\/}(2023)Dong, B{\"u}hler \& Smith]{Dong2023}
{\sc \au{Dong, Wenjing}, \au{B{\"u}hler, Oliver} \& \au{Smith, K.~Shafer}} \yr{2023}  \at{Geostrophic eddies spread near-inertial wave energy to high frequencies}.  \jt{Journal of Physical Oceanography}  \bvol{53}~(5),  \pg{1311 -- 1322}.

\bibitem[Eden {\em et~al.\/}(2019{\natexlab{{\em a\/}}})Eden, Chouksey \& Olbers]{Eden2019a}
{\sc \au{Eden, Carsten}, \au{Chouksey, Manita} \& \au{Olbers, Dirk}} \yr{2019{\natexlab{{\em a\/}}}}  \at{{Gravity wave emission by shear instability}}.  \jt{J. Phys. Oceanogr.}  \bvol{49}~(9),  \pg{2393--2406}.

\bibitem[Eden {\em et~al.\/}(2019{\natexlab{{\em b\/}}})Eden, Pollmann \& Olbers]{Eden2019b}
{\sc \au{Eden, Carsten}, \au{Pollmann, Friederike} \& \au{Olbers, Dirk}} \yr{2019{\natexlab{{\em b\/}}}}  \at{Numerical evaluation of energy transfers in internal gravity wave spectra of the ocean}.  \jt{Journal of Physical Oceanography}  \bvol{49}~(3),  \pg{737 -- 749}.

\bibitem[Eden {\em et~al.\/}(2020)Eden, Pollmann \& Olbers]{Eden2020}
{\sc \au{Eden, Carsten}, \au{Pollmann, Friederike} \& \au{Olbers, Dirk}} \yr{2020}  \at{Towards a global spectral energy budget for internal gravity waves in the ocean}.  \jt{Journal of Physical Oceanography}  \bvol{50}~(4),  \pg{935 -- 944}.

\bibitem[Garrett \& Munk(1972)]{Garrett1972}
{\sc \au{Garrett, Christopher} \& \au{Munk, Walter}} \yr{1972}  \at{Space-time scales of internal waves}.  \jt{Geophysical Fluid Dynamics}  \bvol{3}~(3),  \pg{225--264},  \arxiv{arXiv: https://doi.org/10.1080/03091927208236082}.

\bibitem[Garrett \& Munk(1975)]{Garrett1975}
{\sc \au{Garrett, Christopher} \& \au{Munk, Walter}} \yr{1975}  \at{{Space-time scales of internal waves: A progress report}}.  \jt{J. Geophys. Res.}  \bvol{80}~(3),  \pg{291--297}.

\bibitem[Gregg(1989)]{Gregg1989}
{\sc \au{Gregg, M.~C.}} \yr{1989}  \at{Scaling turbulent dissipation in the thermocline}.  \jt{Journal of Geophysical Research: Oceans}  \bvol{94}~(C7),  \pg{9686--9698},  \arxiv{arXiv: https://agupubs.onlinelibrary.wiley.com/doi/pdf/10.1029/JC094iC07p09686}.

\bibitem[Hasselmann(1966)]{Hasselmann1966}
{\sc \au{Hasselmann, K.}} \yr{1966}  \at{Feynman diagrams and interaction rules of wave-wave scattering processes}.  \jt{Reviews of Geophysics}  \bvol{4}~(1),  \pg{1--32},  \arxiv{arXiv: https://agupubs.onlinelibrary.wiley.com/doi/pdf/10.1029/RG004i001p00001}.

\bibitem[Hasselmann {\em et~al.\/}(1967)Hasselmann, Saffman \& Lighthill]{Hasselmann1967}
{\sc \au{Hasselmann, K.}, \au{Saffman, Philip~Geoffrey} \& \au{Lighthill, Michael~James}} \yr{1967}  \at{Nonlinear interactions treated by the methods of theoretical physics (with application to the generation of waves by wind)}.  \jt{Proceedings of the Royal Society of London. Series A. Mathematical and Physical Sciences}  \bvol{299}~(1456),  \pg{77--103},  \arxiv{arXiv: https://royalsocietypublishing.org/doi/pdf/10.1098/rspa.1967.0124}.

\bibitem[Henyey {\em et~al.\/}(1986)Henyey, Wright \& Flatt{\'e}]{Henyey1986}
{\sc \au{Henyey, Frank~S.}, \au{Wright, Jon} \& \au{Flatt{\'e}, Stanley~M.}} \yr{1986}  \at{Energy and action flow through the internal wave field: An eikonal approach}.  \jt{Journal of Geophysical Research: Oceans}  \bvol{91}~(C7),  \pg{8487--8495},  \arxiv{arXiv: https://agupubs.onlinelibrary.wiley.com/doi/pdf/10.1029/JC091iC07p08487}.

\bibitem[Holloway(1978)]{Holloway1978}
{\sc \au{Holloway, Greg}} \yr{1978}  \at{On the spectral evolution of strongly interacting waves}.  \jt{Geophysical \& Astrophysical Fluid Dynamics}  \bvol{11}~(1),  \pg{271--287},  \arxiv{arXiv: https://doi.org/10.1080/03091927808242670}.

\bibitem[Holloway(1980)]{Holloway1980}
{\sc \au{Holloway, Greg}} \yr{1980}  \at{Oceanic internal waves are not weak waves}.  \jt{Journal of Physical Oceanography}  \bvol{10}~(6),  \pg{906 -- 914}.

\bibitem[Kafiabad {\em et~al.\/}(2019)Kafiabad, Savva \& Vanneste]{Kafiabad2019}
{\sc \au{Kafiabad, Hossein~A}, \au{Savva, Miles~AC} \& \au{Vanneste, Jacques}} \yr{2019}  \at{Diffusion of inertia-gravity waves by geostrophic turbulence}.  \jt{Journal of Fluid Mechanics}  \bvol{869},  \pg{R7}.

\bibitem[Kunze \& {Llewellyn Smith}(2004)]{Kunze2004}
{\sc \au{Kunze, Eric} \& \au{{Llewellyn Smith}, Stefan~G.}} \yr{2004}  \at{{The role of small-scale topography in turbulent mixing of the global ocean}}.  \jt{Oceanography}  \bvol{17}~(SPL.ISS. 1),  \pg{55--64}.

\bibitem[Labarre {\em et~al.\/}(2024{\natexlab{{\em a\/}}})Labarre, Augier, Krstulovic \& Nazarenko]{Labarre2024a}
{\sc \au{Labarre, Vincent}, \au{Augier, Pierre}, \au{Krstulovic, Giorgio} \& \au{Nazarenko, Sergey}} \yr{2024{\natexlab{{\em a\/}}}}  \at{Internal gravity waves in stratified flows with and without vortical modes}.  \jt{Phys. Rev. Fluids}  \bvol{9},  \pg{024604}.

\bibitem[Labarre {\em et~al.\/}(2025)Labarre, Krstulovic \& Nazarenko]{Labarre2025}
{\sc \au{Labarre, Vincent}, \au{Krstulovic, Giorgio} \& \au{Nazarenko, Sergey}} \yr{2025} Wave-kinetic dynamics of forced-dissipated turbulent internal gravity waves,  \arxiv{arXiv: 2407.11469}.

\bibitem[Labarre {\em et~al.\/}(2024{\natexlab{{\em b\/}}})Labarre, Lanchon, Cortet, Krstulovic \& Nazarenko]{Labarre2024b}
{\sc \au{Labarre, Vincent}, \au{Lanchon, Nicolas}, \au{Cortet, Pierre-Philippe}, \au{Krstulovic, Giorgio} \& \au{Nazarenko, Sergey}} \yr{2024{\natexlab{{\em b\/}}}}  \at{On the kinetics of internal gravity waves beyond the hydrostatic regime}.  \jt{Journal of Fluid Mechanics}  \bvol{998},  \pg{A17}.

\bibitem[Lanchon \& Cortet(2023)]{Lanchon2023b}
{\sc \au{Lanchon, Nicolas} \& \au{Cortet, Pierre-Philippe}} \yr{2023}  \at{Energy spectra of nonlocal internal gravity wave turbulence}.  \jt{Phys. Rev. Lett.}  \bvol{131},  \pg{264001}.

\bibitem[Lvov \& Tabak(2004)]{Lvov2004b}
{\sc \au{Lvov, Yuri} \& \au{Tabak, Esteban~G}} \yr{2004}  \at{A hamiltonian formulation for long internal waves}.  \jt{Physica D: Nonlinear Phenomena}  \bvol{195}~(1),  \pg{106--122}.

\bibitem[Lvov {\em et~al.\/}(2010)Lvov, Polzin, Tabak \& Yokoyama]{Lvov2010}
{\sc \au{Lvov, Yuri~V.}, \au{Polzin, Kurt~L.}, \au{Tabak, Esteban~G.} \& \au{Yokoyama, Naoto}} \yr{2010}  \at{Oceanic internal-wave field: Theory of scale-invariant spectra}.  \jt{Journal of Physical Oceanography}  \bvol{40}~(12),  \pg{2605--2623}.

\bibitem[Lvov {\em et~al.\/}(2012)Lvov, Polzin \& Yokoyama]{Lvov2012}
{\sc \au{Lvov, Yuri~V.}, \au{Polzin, Kurt~L.} \& \au{Yokoyama, Naoto}} \yr{2012}  \at{Resonant and near-resonant internal wave interactions}.  \jt{Journal of Physical Oceanography}  \bvol{42}~(5),  \pg{669--691}.

\bibitem[Lvov \& Tabak(2001)]{Lvov2001}
{\sc \au{Lvov, Yuri~V.} \& \au{Tabak, Esteban~G.}} \yr{2001}  \at{Hamiltonian formalism and the garrett-munk spectrum of internal waves in the ocean}.  \jt{Phys. Rev. Lett.}  \bvol{87},  \pg{168--501}.

\bibitem[McComas(1977)]{McComas1977b}
{\sc \au{McComas, C.~H.}} \yr{1977}  \at{Equilibrium mechanisms within the oceanic internal wave field}.  \jt{Journal of Physical Oceanography}  \bvol{7}~(6),  \pg{836--845}.

\bibitem[McComas \& Bretherton(1977{\natexlab{{\em a\/}}})]{McComas1977a}
{\sc \au{McComas, C.~Henry} \& \au{Bretherton, Francis~P.}} \yr{1977{\natexlab{{\em a\/}}}}  \at{Resonant interaction of oceanic internal waves}.  \jt{Journal of Geophysical Research (1896-1977)}  \bvol{82}~(9),  \pg{1397--1412},  \arxiv{arXiv: https://agupubs.onlinelibrary.wiley.com/doi/pdf/10.1029/JC082i009p01397}.

\bibitem[McComas \& Bretherton(1977{\natexlab{{\em b\/}}})]{McComas1977}
{\sc \au{McComas, C.~H.} \& \au{Bretherton, F.~P.}} \yr{1977{\natexlab{{\em b\/}}}}  \at{{Resonant interaction of oceanic internal waves.}}  \jt{J. Geophys. Res.}  \bvol{82}~(9 , Mar.20, 1977),  \pg{1397--1412}.

\bibitem[McComas \& M{\"u}ller(1981{\natexlab{{\em a\/}}})]{McComas1981a}
{\sc \au{McComas, C.~Henry} \& \au{M{\"u}ller, Peter}} \yr{1981{\natexlab{{\em a\/}}}}  \at{The dynamic balance of internal waves}.  \jt{Journal of Physical Oceanography}  \bvol{11}~(7),  \pg{970--986}.

\bibitem[McComas \& M{\"u}ller(1981{\natexlab{{\em b\/}}})]{McComas1981b}
{\sc \au{McComas, C.~Henry} \& \au{M{\"u}ller, Peter}} \yr{1981{\natexlab{{\em b\/}}}}  \at{Time scales of resonant interactions among oceanic internal waves}.  \jt{Journal of Physical Oceanography}  \bvol{11}~(2),  \pg{139--147}.

\bibitem[M{\"u}ller \& Olbers(1975)]{Muller1975}
{\sc \au{M{\"u}ller, Peter} \& \au{Olbers, Dirk~J.}} \yr{1975}  \at{On the dynamics of internal waves in the deep ocean}.  \jt{Journal of Geophysical Research (1896-1977)}  \bvol{80}~(27),  \pg{3848--3860},  \arxiv{arXiv: https://agupubs.onlinelibrary.wiley.com/doi/pdf/10.1029/JC080i027p03848}.

\bibitem[Nazarenko(2011)]{Nazarenko2011}
{\sc \au{Nazarenko, S}} \yr{2011} {\em Wave turbulence\/}.  \publ{Springer}.

\bibitem[Olbers(1974)]{Olbers1974}
{\sc \au{Olbers, Dirk~Jurgen.}} \yr{1974} {\em On the energy balance of small-scale internal waves in the deep-sea\/}.  \publ{G. M. L. Wittenborn Hamburg}.

\bibitem[Olbers(1976)]{Olbers1976}
{\sc \au{Olbers, Dirk~J.}} \yr{1976}  \at{Nonlinear energy transfer and the energy balance of the internal wave field in the deep ocean}.  \jt{Journal of Fluid Mechanics}  \bvol{74}~(2),  \pg{375--399}.

\bibitem[Pan {\em et~al.\/}(2020)Pan, Arbic, Nelson, Menemenlis, Peltier, Xu \& Li]{Pan2020}
{\sc \au{Pan, Yulin}, \au{Arbic, Brian~K.}, \au{Nelson, Arin~D.}, \au{Menemenlis, Dimitris}, \au{Peltier, W.~R.}, \au{Xu, Wentao} \& \au{Li, Ye}} \yr{2020}  \at{Numerical investigation of mechanisms underlying oceanic internal gravity wave power-law spectra}.  \jt{Journal of Physical Oceanography}  \bvol{50}~(9),  \pg{2713 -- 2733}.

\bibitem[Polzin \& Lvov(2011)]{Polzin2011}
{\sc \au{Polzin, K.~L.} \& \au{Lvov, Y.~V.}} \yr{2011}  \at{Toward regional characterizations of the oceanic internal wavefield}.  \jt{Reviews of Geophysics}  \bvol{49}~(4),  \arxiv{arXiv: https://agupubs.onlinelibrary.wiley.com/doi/pdf/10.1029/2010RG000329}.

\bibitem[Polzin {\em et~al.\/}(2014)Polzin, Naveira~Garabato, Huussen, Sloyan \& Waterman]{Polzin2014}
{\sc \au{Polzin, Kurt~L.}, \au{Naveira~Garabato, Alberto~C.}, \au{Huussen, Tycho~N.}, \au{Sloyan, Bernadette~M.} \& \au{Waterman, Stephanie}} \yr{2014}  \at{Finescale parameterizations of turbulent dissipation}.  \jt{Journal of Geophysical Research: Oceans}  \bvol{119}~(2),  \pg{1383--1419},  \arxiv{arXiv: https://agupubs.onlinelibrary.wiley.com/doi/pdf/10.1002/2013JC008979}.

\bibitem[Polzin {\em et~al.\/}(1995)Polzin, Toole \& Schmitt]{Polzin1995}
{\sc \au{Polzin, Kurt~L.}, \au{Toole, John~M.} \& \au{Schmitt, Raymond~W.}} \yr{1995}  \at{Finescale parameterizations of turbulent dissipation}.  \jt{Journal of Physical Oceanography}  \bvol{25}~(3),  \pg{306 -- 328}.

\bibitem[Savva {\em et~al.\/}(2021)Savva, Kafiabad \& Vanneste]{Savva2021}
{\sc \au{Savva, M.~A.C.}, \au{Kafiabad, H.~A.} \& \au{Vanneste, J.}} \yr{2021}  \at{{Inertia-gravity-wave scattering by three-dimensional geostrophic turbulence}}.  \jt{J. Fluid Mech.}  \bvol{916},  \pg{1--30}.

\bibitem[Skitka {\em et~al.\/}(2024)Skitka, Arbic, Thakur, Menemenlis, Peltier, Pan, Momeni \& Ma]{Skitka2024a}
{\sc \au{Skitka, Joseph}, \au{Arbic, Brian~K.}, \au{Thakur, Ritabrata}, \au{Menemenlis, Dimitris}, \au{Peltier, William~R.}, \au{Pan, Yulin}, \au{Momeni, Kayhan} \& \au{Ma, Yuchen}} \yr{2024}  \at{Probing the nonlinear interactions of supertidal internal waves using a high-resolution regional ocean model}.  \jt{Journal of Physical Oceanography}  \bvol{54}~(2),  \pg{399 -- 425}.

\bibitem[Wu \& Pan(2023)]{Wu2023b}
{\sc \au{Wu, Yue} \& \au{Pan, Yulin}} \yr{2023}  \at{Energy cascade in the garrett--munk spectrum of internal gravity waves}.  \jt{Journal of Fluid Mechanics}  \bvol{975},  \pg{A11}.

\bibitem[Zakharov {\em et~al.\/}(1992)Zakharov, Lvov \& Falkovich]{Zakharov1992}
{\sc \au{Zakharov, V.~E.}, \au{Lvov, V.~S.} \& \au{Falkovich, G.}} \yr{1992} {\em Kolmogorov Spectra of Turbulence\/}.  \publ{Springer-Verlag}.

\end{thebibliography}

\end{document}